\documentclass[letterpaper, conference]{IEEEtran}
\IEEEoverridecommandlockouts

\usepackage{cite}
\usepackage{amsmath}
\usepackage{amssymb}
\usepackage{amsfonts}
\usepackage{algorithmic}
\usepackage{graphicx}
\usepackage{textcomp}
\usepackage{xcolor}
\usepackage{tikz}
\usepackage{graphics}
\usepackage{epsfig}
\usepackage{times}
\usepackage{bm}
\usepackage{psfrag}
\usepackage{mathtools}
\usepackage{picture}
\usepackage{xcolor}
\usepackage[colorinlistoftodos]{todonotes}
\usepackage{float}
\usepackage{verbatimbox}
\usepackage{listings}
\usepackage{enumerate}
\usepackage{bigints}
\usepackage{ntheorem}
\usepackage[utf8]{inputenc}
\usepackage[english]{babel}

\makeatletter
\renewtheoremstyle{plain}
{\item{\theorem@headerfont ##1\ ##2\theorem@separator}~}
{\item{\theorem@headerfont ##1\ ##2\ (##3)\theorem@separator}~}
\theoremseparator{:}
\makeatother

\newtheorem{problem}{Problem}
\newtheorem{assumption}{Assumption}
\newtheorem{remark}{Remark}

\newcommand\at[2]{\left.#1\right|_{#2}}
\newcommand\copyrighttext{
\footnotesize
\textit{\textcopyright{}~2021 IEEE.  Personal use of this material is permitted.  Permission from IEEE must be obtained for all other uses, in any current or future media, including reprinting/republishing this material for advertising or promotional purposes, creating new collective works, for resale or redistribution to servers or lists, or reuse of any copyrighted component of this work in other works.}
}
\newcommand\copyrightnotice{
\begin{tikzpicture}[remember picture,overlay]
\node[anchor=south,yshift=20pt] at (current page.south) {\parbox{\dimexpr\textwidth-\fboxsep-\fboxrule\relax}{\copyrighttext}};
\end{tikzpicture}
}

\def\BibTeX{{\rm B\kern-.05em{\sc i\kern-.025em b}\kern-.08em
    T\kern-.1667em\lower.7ex\hbox{E}\kern-.125emX}}
\begin{document}

\title{Low-Input Accurate Periodic Motion of an Underactuated Mechanism: Mass Distribution and Nonlinear Spring Shaping
}
\author{
\IEEEauthorblockN{Andrea Tilli, Elena Ruggiano, Alessandro Bosso, and Alessandro Samor\`i}
\IEEEauthorblockA{\textit{Department of Electrical, Electronic, and Information Engineering (DEI)}\\
\textit{University of Bologna, 40136, Bologna, Italy}\\
Email:{\tt\{andrea.tilli,elena.ruggiano,alessandro.bosso,alessandro.samori4\}@unibo.it}}
}
\maketitle
\copyrightnotice

\begin{abstract}
This work presents a control-oriented structural design approach for a 2-DOF underactuated mechanical system, with the purpose of generating an optimal oscillatory behavior of the end-effector.
To achieve the desired periodic motion, we propose to adjust the dynamic response of the mechanism by selecting its mass distribution and the characteristic of a nonlinear spring.
In particular, we introduce a two-step optimization strategy to shape the system's zero dynamics, obtained via input-output linearization.
The first part of the procedure aims to minimize the root-mean-square value of the input torque by optimizing the mechanism's mass distribution.
In this context, we show that a perfect matching with the desired trajectory can be reached by assuming the ability to design an arbitrary shape of the system's elastic properties.
Then, in order to favor a simpler physical implementation of the structure, we dedicate the second optimization step to the piecewise linear approximation of the previously defined stiffness characteristic.
The proposed procedure is finally tested in detailed numerical simulations, confirming its effectiveness in generating a complex and efficient periodic motion.
\end{abstract}


\section{Introduction}
In recent years, the design of high-performance and sustainable mechanical systems has become an ever-growing challenge.
This topic is especially relevant in industrial applications, where it is crucial to obtain an accurate motion of the end-effector while minimizing as much as possible the realization costs and the energy consumption.
However, in such a context, a prevalent practice consists of making the engineering design a sequential procedure, so that the control problem is approached only when the structure is fully established.
Inevitably, this approach leads to oversized components and a high impact on costs and efficiency.
\par Due to the severe disadvantages of sequential strategies, this work focuses on a more integrated approach, where the control requirements are evaluated at the early design stage.
In particular, our objective is to achieve high-performance motion planning for multibody systems through structural optimization.
This challenging topic has received, in recent years, widespread interest by the scientific community.
We refer to \cite{tromme2018system, ma2006multidomain} for detailed overviews regarding structural and topological optimization of multibody systems.
\par Within the domain of multibody systems, we specialize the analysis of this work to underactuated Euler-Lagrange systems, i.e., mechanisms having fewer control inputs than the Degrees of Freedom (DOF).
The specific properties of these systems and the related control techniques are examined, e.g., in \cite{liu2013survey} and \cite{de2002underactuated}.
These mechanisms are often desirable because they involve a reduced overall actuation.
However, not all DOF can be controlled simultaneously; as a consequence, such systems present internal dynamics whose properties cannot be entirely influenced by the control inputs.
Therefore, structural design becomes an essential tool to impose a desirable behavior of the mechanism. 
In this respect, \cite{seifried2014dynamics} poses the attention to shaping the system's zero dynamics, i.e., the internal dynamics in conditions of perfect reference tracking.
There, the author defines an integrated structure-control optimization procedure aimed at obtaining a minimum phase system, i.e., stable zero dynamics.
In \cite{bastos2019synergistic}, an integrated method is proposed to minimize both the feedforward control input and the internal dynamics motion of underactuated manipulators.
\par In this paper, we present an optimal design strategy for a 2-DOF underactuated mechanism, with the purpose of following a periodic reference trajectory.
The proposed approach consists of finding a minimum control effort by optimizing some relevant structural parameters so that the resulting system oscillates according to a desired periodic reference trajectory.
A closely related problem has been considered in \cite{tilli2020periodic}.
However, here we do not focus on constructing a comprehensive control law stabilizing the system on the desired orbit.
Instead, we perform input-output linearization and we approach the structural design in conditions of perfect output tracking.
One of the main contributions of this paper, derived from the formalism in \cite{shiriaev2005constructive}, is showing the connection between the system's stiffness and its resulting zero dynamics.
In particular, we point out that the presence of at least one customizable compliant joint is sufficient to shape the zero dynamics arbitrarily.
Then, inspired by the possibility of realizing nonlinear springs \cite{friswell2012vibration} and by the solutions found in \cite{radaelli2017static, yao2018novel}, we employ a piecewise linear characteristic as a means to approximate arbitrary stiffness models.
\par The proposed optimization approach is based on the following two-step solution.
Firstly, we optimize the system's mass distribution to minimize the root-mean-square (RMS) value of the input torque required to maintain the system on the desired trajectory.
Contextually, we also derive an ideal nonlinear stiffness characteristic associated with a compliant joint.
Then, the second step consists of designing a piecewise linear spring that closely matches the ideal one. 
\par The paper is organized as follows.
In Section \ref{sec:cntr-struc_design}, we formalize the problem statement and derive a minimal form model useful for the proposed solution.
In Section \ref{sec:cntr-struc_design}, we also derive the explicit form of the zero dynamics and highlight how the system's elasticity properties influence it.
In Section \ref{sec:two-step_sol}, the two-step optimization strategy is presented.
Then, in Section \ref{sec:num_results}, numerical simulation results are shown for validation, while some final considerations are reported in Section \ref{sec:conclusions}.

\section{Control-oriented structural design for periodic motion optimization}\label{sec:cntr-struc_design}

\subsection{Problem Statement}

\begin{figure}
\centering
\psfragscanon
	
\vspace{3pt}
	
\psfrag{a} [B][B][0.7][0]{$q_1$}
\psfrag{b} [B][B][0.7][0]{$q_2$}
\psfrag{c} [B][B][0.7][0]{$q_3$}
\psfrag{d} [B][B][0.7][0]{$q_4$}
\psfrag{l} [B][B][0.7][0]{$\mathcal{L}_1$}
\psfrag{m} [B][B][0.7][0]{$\mathcal{L}_2$}
\psfrag{n} [B][B][0.7][0]{$\mathcal{L}_3$}
\psfrag{o} [B][B][0.7][0]{$\mathcal{L}_4$}
\psfrag{e} [B][B][0.7][0]{$\theta$}
\psfrag{f} [B][B][0.7][0]{$\delta_i$}
\psfrag{p} [B][B][0.7][0]{$\mathcal{J}_1$}
\psfrag{q} [B][B][0.7][0]{$\mathcal{J}_2$}
\psfrag{r} [B][B][0.7][0]{$\mathcal{J}_3$}
\psfrag{s} [B][B][0.7][0]{$\mathcal{J}_4$}
\psfrag{t} [B][B][0.7][0]{$\mathcal{J}_5$}
\psfrag{i} [B][B][0.7][0]{$m_{A_i}$}
\psfrag{u} [B][B][0.7][0]{$u$}
\psfrag{v} [B][B][0.7][0]{$\mathcal{J}_i$}
\psfrag{z} [B][B][0.7][0]{$\mathcal{J}_{i+1}$}
\psfrag{h} [B][B][0.7][0]{$h(q) = \begin{pmatrix}x\\ y\end{pmatrix}$}
\psfrag{x} [B][B][0.7][0]{$e_x$}
\psfrag{y} [B][B][0.7][0]{$e_y$}
\psfrag{g} [B][B][0.7][0]{g}
\includegraphics[clip = true, width = 0.45\textwidth]{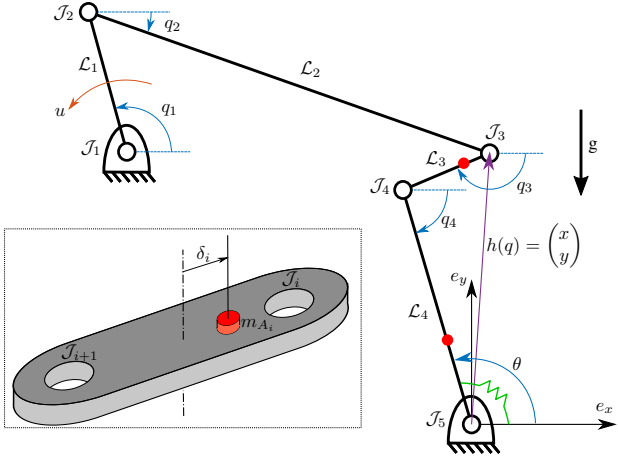}	
\caption{Geometric scheme of the 2-DOF mechanism used as a case of study. $\mathcal{L}_i, i \in \{1, \hdots, 4\}$ indicate the links, while $\mathcal{J}_i, i \in \{1, \hdots, 5\}$ denote the joints. The green symbol indicates the presence of a torsional spring in joint $\mathcal{J}_5$. The joint coordinates are positive when associated with a counterclockwise rotation (hence $q_2$, $q_3$, $q_4$ are negative in the figure), while the pair $(e_x, e_y)$ is the reference frame of the workspace. In such a frame, the position $h$ of the end-effector is indicated in violet. The bottom-left box shows in detail the structure of a link equipped with an additional mass.}
\label{fig:mechanism}
\end{figure}

We begin by introducing a precise mathematical framework for the proposed optimization strategy.
Although our approach is compatible, in principle, with any underactuated 2-DOF Euler-Lagrange system, we specialize the discussion to the mechanism in Fig. \ref{fig:mechanism}.
This structure, employed in \cite{tilli2020periodic} as a case of study for a closely related structure-control design, is a modified 4-bar-linkage where the rocker is replaced with two links.
In particular, the purpose of the mechanism is to achieve a periodic horizontal motion of the end-effector, given by the position of joint $\mathcal{J}_3$ in the $(e_x, e_y)$ frame, applying a torque $u$ to joint $\mathcal{J}_1$ (while all other joints are passive).
Let $q = (q_1 \ q_2 \ q_3 \ q_4)^\top \in \mathbb{R}^4$ denote the angles associated with joints $\mathcal{J}_i$, $i \in \{1, \ldots, 4\}$.
In order to appropriately optimize the behavior of the mechanism, not only the input $u$ but also some structural elements are available as degrees of freedom.
In particular, we suppose that the system can be shaped through the following features:
\begin{itemize}
\item two masses attached to links $\mathcal{L}_3$ and $\mathcal{L}_4$ (in red in Fig. \ref{fig:mechanism}), with tunable values $m_{A_i}$ and offsets $\delta_{i}$,  $i \in \{3, 4\}$.
The masses can be placed on the longitudinal axis of the link and $\delta_{i}$ denotes the offset between the mass position and the geometric center of the link.
Such offset is positive when the mass is closer to joint $\mathcal{J}_i$.
The design parameters are thus collected as:
\begin{equation}
\label{eq:vector_p1}
p_m = (m_{A_3} \ m_{A_4} \ \delta_3 \ \delta_4)^\top;
\end{equation}
\item a torsional spring placed in joint $\mathcal{J}_5$ (in green in Fig. \ref{fig:mechanism}), with a characteristic that can be customized according to a desired (finite-dimensional) class of functions $\mathcal{S}$.
In particular, we employ a model of the form $\tau_s = S(p_s, \theta(q))$, where $\tau_s$ indicates the spring output torque and $\theta(q) = q_4 + \pi$ is the angular displacement.
Finally, it holds $S(p_s, \cdot) \in \mathcal{S}$, where $p_s \in \mathbb{R}^d$ is used to parameterize the set $\mathcal{S}$.
The choice of $\mathcal{S}$ will be specified in section \ref{sec:stiff_approx}.
\end{itemize}
In view of the above definitions, the model of the mechanism is given by the following differential algebraic equation (DAE): 
\begin{equation}
\label{eq:model_q_coord}
\begin{split}
&M_{q} (p_m, q) \ddot{q} \hspace{-1pt} + \hspace{-1pt} C_{q} (p_m, q, \dot{q}) \dot{q} \hspace{-1pt} + \hspace{-1pt} G_{q} (p_m, p_s, q) \hspace{-1pt} = \hspace{-1pt} B_{q} u \hspace{-1pt} + \hspace{-1pt} u_{\phi}(q, \hspace{-1pt} \lambda) \\
&\phi(q) = 0, \qquad u_{\phi}(q, \lambda) = \left[\frac{\partial \phi}{\partial q}(q)\right]^\top \lambda,
\end{split}
\end{equation}
where $M_{q} (p_m, q) \in \mathbb{R}^{4 \times 4}$ is the inertia matrix, $C_{q}(p_m, q, \dot{q}) \in \mathbb{R}^{4 \times 4}$ denotes the centrifugal and Coriolis effects, $G_{q} (p_m, p_s, q) \in \mathbb{R}^4$ is the term accounting for the potential energy contributions (gravity and elasticity), while $B_{q} = (1 \ 0 \ 0 \ 0)^\top$ is the input matrix.
In addition, $\lambda \in \mathbb{R}^2$ are the Lagrange multipliers associated with the kinematic loop constraints $\phi(q) = 0$, where $\phi: \mathbb{R}^4 \rightarrow \mathbb{R}^2$ is a smooth map.
For convenience in the design, we let $p_m \in \mathcal{P}_m$ and $p_s \in \mathcal{P}_s$, where $\mathcal{P}_m \subset\mathbb{R}^4$ and $\mathcal{P}_s \subset \mathbb{R}^d$ are compact and convex sets.
Notice that no friction nor damping effects are considered. 
\par In the following, we consider an output for control $\chi$ given by the workspace coordinates of the end-effector (see Fig. \ref{fig:mechanism}):
\begin{equation}
\label{eq:direct_kin}
\chi = h(q) = \begin{pmatrix}
h_1(q) \\
h_2(q)
\end{pmatrix} = 
\begin{pmatrix}
x \\
y
\end{pmatrix},
\end{equation}
where $h: \mathbb{R}^4 \rightarrow \mathbb{R}^2$ is a smooth map representing the direct kinematics of the system.
Notice that the kinematic properties of the mechanism are not influenced by $p_m$ and $p_s$.
The design objectives are formally collected in the following statement.
\begin{problem}\label{prob1}
Consider a periodic reference in the workspace, with period $T_r > 0$:
\begin{equation}
\label{eq:ref_traj}
r(t) = \begin{pmatrix}
r_x(t) \\
\bar{r}_y
\end{pmatrix}
\end{equation}
with $\bar r_y$ a constant scalar and $r_x(\cdot) \in \mathcal{C}^2$.
Design:
\begin{itemize}
\item a set of parameters $\bar{p}_m \in \mathcal{P}_m$;
\item a set of parameters $\bar{p}_s \in \mathcal{P}_s$;
\item an input torque of the form
\begin{equation}
\label{eq:generic_ctrl_u}
\bar{u} = \mu(\bar{p}_m, \bar{p}_s, q,\dot q),
\end{equation} 
where $\mu$ is an appropriate map;
\end{itemize}
such that the following properties are satisfied:
\begin{itemize}
\item there exists a solution $(\bar{q}(t), \dot{\bar{q}}(t))$ of the closed-loop system \eqref{eq:model_q_coord}, \eqref{eq:generic_ctrl_u} that is periodic, with period $T_q$, and satisfies $(\bar{x}(0), \dot{\bar{x}}(0), \bar{y}(0), \dot{\bar{y}}(0)) = (r_x(0), \dot{r}_x(0), \bar{r}_y, 0)$, where $(\bar x(t) \ \bar{y}(t))^\top = h(\bar q(t))$;
\item $(\bar{q}(t), \dot{\bar{q}}(t))$ is such that the set $\Gamma_\chi = \{\xi \in \mathbb{R}^4: \xi = (\bar{x}(t), \dot{\bar{x}}(t), \bar{y}(t), \dot{\bar{y}}(t)), t \in [0, T_q] \}$ approximates, as much as possible, the set $\Gamma_r = \{\xi \in \mathbb{R}^4: \xi = (r_x(t), \dot{r}_x(t), \bar{r}_y, 0), t \in [0, T_r] \}$;
\item the RMS value of $\bar{u}$:
\begin{equation}\label{eq:problem_cost_function}
\bar{u}_{\text{RMS}}(\bar{p}_m, \bar{p}_s) = \sqrt{\frac{1}{T_q}\int_{0}^{T_q}\mu^2(\bar{p}_m, \bar{p}_s, \bar{q}(s),\dot{\bar{q}}(s)) \ ds}
\end{equation}
is minimized.
\end{itemize}
\end{problem}
\begin{remark}
\label{rem1}
We highlight that the feedback law \eqref{eq:generic_ctrl_u} should not be intended as a full control solution ensuring convergence of the trajectories $(q(t), \dot{q}(t))$ to $(\bar{q}(t), \dot{\bar{q}}(t))$ or to its orbit, given by $\Gamma_q = \{(\xi_1, \xi_2) \in \mathbb{R}^4\times\mathbb{R}^4: \xi_1 = \bar{q}(t), \xi_2 = \dot{\bar{q}}(t), t \in [0, T_q] \}$.
In order to ensure this feature, it is necessary to design an additional input law $u_{\textup{st}}$ such that the control law $u = \bar{u} + u_{\textup{st}}$
renders either $(\bar{q}(t), \dot{\bar{q}}(t))$ or $\Gamma_q$ attractive from a given set of initial conditions, with appropriate stability properties.
We leave this problem out of the purpose of this work, while we refer to \cite{tilli2020periodic} for a solution based on the orbital stabilization techniques found in \cite{shiriaev2010transverse}. 
\end{remark}
In a fully actuated scenario (i.e., if another joint were actuated), the computation of the input torque $\bar{u}$ could be achieved, e.g., via full-state feedback linearization.
However, in the considered underactuated scenario, we lack a complete control authority on both output coordinates $x$ and $y$.
Therefore, the input $\bar{u}$ cannot be designed to exactly impose the reference trajectory $r$, unless such a trajectory is compatible with the behavior of the resulting internal dynamics.
The main intuition of this paper, formally described in the following, is to shape the input torque to keep the end-effector precisely on the reference height $\bar r_y$, while shaping the spring characteristic $S$ so that the orbit of $(x(t), \dot{x}(t))$ closely matches the one associated with $(r_x(t), \dot{r}_x(t))$.
In this perspective, the selection of $p_m$ is aimed at minimizing $\bar{u}_{\text{RMS}}$ without affecting $(x(t), \dot{x}(t))$.

\subsection{Representation of the Mechanism in Minimal Form}\label{sec:min_form}
The procedure to convert model \eqref{eq:model_q_coord} to the minimal form model is quite standard and is presented in detail in \cite[Section 2.1.5]{seifried2014dynamics}.
This procedure consists of mapping the coordinate vector $q$ in some independent coordinates $q_\text{i} \in \mathbb{R}^2$ and some dependent coordinates $q_\text{d} \in \mathbb{R}^2$.
In this work, we follow the same arguments in \cite{tilli2020periodic} to specialize the general approach to the selection $q_{\text{i}} = \chi$, so that we can express the model in the workspace coordinates $(x, y)$.
\begin{assumption}\label{hyp:change_of_coordinates}
The maps $\phi$ and $h$ are such that:
\begin{itemize}
\item 
there exists a differentiable map $q_r: \mathbb{R}^2 \to \mathbb{R}^4$ such that $h(q_r(r(t)))=r(t)$ and $\phi(q_r(r(t))) = 0$, for all $t \in [0, T_r]$;
\item  
there exist open sets $\mathcal{U}, \mathcal{V} \subset \mathbb{R}^4$ and a diffeomorphism $\rho: \mathcal{U} \to \mathcal{V}$, satisfying $q = \rho(\chi, \psi)$ (for some $\psi \in \mathbb{R}^2$) and $q_r(r(t)) \in \mathcal{V}$, for all $t \in [0, T_r]$;
\item 
the matrix
\begin{equation}
\frac{\partial \phi}{\partial q}\frac{\partial \rho}{\partial \psi}(\chi, \psi)
\end{equation}
is non-singular in $(\chi, \psi) = \rho^{-1}(q_r(r(t))$, $\forall t \in [0, T_r]$.
\end{itemize}
\end{assumption}
This way, it is straightforward to show that there exists an open set, $\mathcal{N} \subset \mathbb{R}^2$, such that $r(t) \in \mathcal{N}$, for all $t \in [0, T_r]$, and such that system \eqref{eq:model_q_coord} can be rewritten as
\begin{equation}\label{eq:chi_system}
M(p_m, \chi)\ddot{\chi} + C(p_m, \chi, \dot{\chi})\dot{\chi} + G(p_m, p_s, \chi) = B(\chi)u, \;\; \chi \in \mathcal{N}
\end{equation}
where we express the data as follows:
\begin{itemize}
\item inertia matrix:
\begin{equation}
M(p_m, \chi) = \begin{bmatrix}
M_{11}(p_m, \chi)  &M_{12}(p_m, \chi) \\
M_{12}(p_m, \chi)  &M_{22}(p_m, \chi)
\end{bmatrix};
\end{equation}
\item Coriolis and centrifugal torques:  $C(p_m, \chi, \dot{\chi})\dot{\chi} = [C_1(p_m, \chi, \dot \chi) \quad C_2(p_m, \chi, \dot \chi)]^\top$, where:
\begin{equation}\label{eq:C_terms}
\begin{split}
C_1 &= \frac{1}{2} \frac{\partial M_{11}}{\partial x} {\dot x}^2 + \frac{\partial M_{11}}{\partial y} \dot x \dot y + \left( \frac{\partial M_{12}}{\partial y} - \frac{1}{2} \frac{\partial M_{22}}{\partial x} \right) {\dot y}^2, \\
C_2 &= \left( \frac{\partial M_{12}}{\partial x} - \frac{1}{2} \frac{\partial M_{11}}{\partial y} \right) {\dot x}^2 + \frac{\partial M_{22}}{\partial x} \dot x \dot y + \frac{1}{2} \frac{\partial M_{22}}{\partial y} {\dot y}^2;
\end{split}
\end{equation}
\item potential energy derivative, accounting for the gravity and the elasticity contributions: $G(p_m, p_s, \chi) = [G_1(p_m, p_s, \chi) \quad G_2(p_m, p_s, \chi)]^\top$:
\begin{equation}
\label{eq:G1_G2}
\hspace{-5pt} G_1 \hspace{-2pt} = \hspace{-2pt} S\left(p_s, \theta\right) \hspace{-2pt} \frac{\partial q_4}{\partial x} \hspace{-1pt} + \hspace{-1pt} \frac{\partial U_\text{g}}{\partial x},  \,\,
G_2 \hspace{-2pt} = S\left(p_s, \theta\right) \hspace{-2pt} \frac{\partial q_4}{\partial y} \hspace{-1pt} + \hspace{-1pt} \frac{\partial U_\text{g}}{\partial y}, \hspace{-7pt}
\end{equation}
where $U_\text{g}(p_m, \chi)$ denotes the gravity potential energy. 
Notice that $\theta$, previously defined as $\theta = \pi + q_4$, can be equivalently expressed, exploiting the inverse kinematics ($q=h^{-1}(\chi)$), as a function of the workspace coordinates: $\theta = \eta(x,y)$.
\item the input distribution vector $B(\chi) = [B_1(\chi) \quad B_2(\chi)]^\top$, where $B_1 = \partial q_1/\partial x$ and $B_2 = \partial q_1/\partial y$.
\end{itemize}

\subsection{Input-Output Feedback Linearization}\label{sec:feedback_lin}
As anticipated, our strategy consists of employing the input torque to maintain the $y$ coordinate on the reference $\bar{r}_y$.
The resulting zero dynamics, i.e., the residual system obtained under conditions of perfect tracking for the height $y$, can be written in terms of only $x$ and $\dot{x}$.
Such dynamics are crucial for the subsequent design, since the available structural parameters can be adjusted to match the behavior of $x$ with the reference $r_x$.
We now illustrate how the zero dynamics are computed via an input-output linearizing control law.
\par Taking advantage of the explicit expressions of $M$, $C \dot{\chi}$, $G$, and $B$, and inverting the inertia matrix, we can rewrite system \eqref{eq:chi_system} as follows: 
\begin{equation}\label{eq:f_g_system}
\begin{split}
\ddot x &= f_x(p_m, p_s, x, \dot x, y, \dot y) + g_x(p_m, x, y) u\\ 
\ddot y &= f_y(p_m, p_s, x, \dot x, y, \dot y) + g_y(p_m, x, y) u
\end{split}
\end{equation}
where: 
\begin{equation}
\label{eq_f_g_explicit}
\begin{split}
f_x &= \frac{-M_{22}(C_1 + G_1) + M_{12}(C_2 + G_2)}{M_{11}M_{22} - M_{12}^2}\\
g_x &= \frac{M_{22}B_1 - M_{12}B_2}{M_{11}M_{22} - M_{12}^2} \\
f_y &= \frac{M_{12}(C_1 + G_1) - M_{11}(C_2 + G_2)}{M_{11}M_{22} - M_{12}^2} \\
g_y &= - \frac{M_{12}B_1 - M_{11}B_2}{M_{11}M_{22} - M_{12}^2}.
\end{split}
\end{equation}
For the feasibility of the proposed approach, consider the following Assumption.
\begin{assumption}
For all $p_m \in \mathcal{P}_m$ and all $\chi \in \mathcal{N}$, it holds $g_y(p_m, \chi) \neq 0$.
\end{assumption}
This way, we can apply to system \eqref{eq:f_g_system} a torque of the form
\begin{equation}
u = -\frac{f_y(p_m, p_s, x, \dot x, y, \dot y)}{g_y(p_m, x, y)} + \frac{v}{g_y(p_m, x, y)},
\end{equation}
for a generic input $v$, which can be employed in the perspective of Remark \ref{rem1} to define a controller for asymptotic tracking or orbital stabilization.
We can rewrite system \eqref{eq:f_g_system} as
\begin{equation}\label{eq:f_g_inverse}
\begin{split}
\ddot x =&\; f_x(p_m, p_s, x, \dot x, y, \dot y)\\
&\qquad -\frac{g_x(p_m, x, y)}{g_y(p_m, x, y)}[f_y(p_m, p_s, x, \dot x, y, \dot y) + v] \\ 
\ddot y =&\; v,
\end{split}
\end{equation}
therefore the zero dynamics are computed by imposing $y = \bar{r}_y$ and $v = 0$, for all $t$, leading to:
\begin{equation}\label{eq:zero_dyn_f_g}
\ddot x \hspace{-2pt} = \hspace{-2pt} f_x(p_m, p_s, x, \dot x, \bar r_y, \hspace{-1pt} 0)  -  \frac{g_x(p_m, x, \bar r_y)}{g_y(p_m, x, \bar r_y)} \hspace{-1pt} f_y(p_m, p_s, x, \dot x, \bar r_y, \hspace{-1pt} 0).
\end{equation}
Let $\bar M, \bar G$ and $\bar B$ be the respective maps evaluated in $y = \bar r_y$:
\begin{equation}
\begin{split}
\bar M(p_m, x) &= M(p_m, x, \bar r_y), \\
\bar G(p_m, p_s, x) &= G(p_m, p_s, x, \bar r_y), \\
\bar B(x) &= B(x, \bar r_y),
\end{split}
\end{equation}
and let $\bar C$ be $C$ computed for $y = \bar r_y$ and for $\dot y = 0$, i.e., $\bar C(p_m,x,\dot x) = C(p_m, x,\dot x, \bar r_y, 0)$.
Notice that, having set $\dot y = 0$, \eqref{eq:C_terms} can be simplified as follows: 
\begin{equation}\label{eq:C1_C2_simply}
\begin{split}
\bar C_1(p_m, x, \dot x) &= C_1(p_m, x,\dot x, \bar r_y, 0) = \frac{1}{2} \at{\frac{\partial M_{11}}{\partial x}}{y = \bar{r}_y} {\dot x}^2,  \\
\bar C_2(p_m, x, \dot x) &= C_2(p_m, x,\dot x, \bar r_y, 0) = \\
&\hspace{1.2cm} = \at{\left( \frac{\partial M_{12}}{\partial x} - \frac{1}{2} \frac{\partial M_{11}}{\partial y} \right)}{y = \bar{r}_y} {\dot x}^2.
\end{split}
\end{equation}
Due to \eqref{eq_f_g_explicit}, system \eqref{eq:zero_dyn_f_g} is equivalent to: 
\begin{equation}\label{eq:zero_dyn_M_C_G}
(\bar M_{11} \bar B_2 - \bar M_{12} \bar B_1) \ddot x = \bar B_1(\bar C_2 + \bar G_2) - \bar B_2 (\bar C_1 + \bar G_1).
\end{equation}
while the residual forcing input torque $u_x(p_m, p_s, x, \dot x)$ can be expressed as:
\begin{equation}\label{eq:u_ideal}
u_x \hspace{-3pt} = \hspace{-2pt} - \frac{f_y(p_m, p_s, x, \dot x, \bar r_y, 0)}{g_y(p_m, x, \bar r_y)} \hspace{-2pt} = \hspace{-2pt} \frac{\bar M_{12} \hspace{-1pt} (\bar C_1 \hspace{-2pt} + \hspace{-2pt} \bar G_1) \hspace{-2pt} - \hspace{-2pt} \bar M_{11} \hspace{-1pt}(\bar C_2 \hspace{-2pt} + \hspace{-2pt} \bar G_2)}{\bar M_{12}\bar B_1 - \bar M_{11}\bar B_2} \hspace{-1pt}.
\end{equation}
\par Finally, combining \eqref{eq:C1_C2_simply} and \eqref{eq:zero_dyn_M_C_G}, we can rewrite the zero dynamics as follows: 
\begin{equation}\label{eq:zero_dyn_alpha_beta_gamma}
\alpha(p_m,x) \ddot x + \beta(p_m,x) \dot x^2 + \gamma(p_m, p_s, x) = 0,
\end{equation}
where
\begin{equation}\label{eq:alpha_beta_gamma}
\begin{split}
\alpha(p_m, x) &= \bar M_{11}(p_m,x) \bar B_2(x) - \bar M_{12}(p_m,x) \bar B_1(x) \neq 0, \\
\beta(p_m,x) &= - \bar B_1(x) \at{\left(\frac{\partial M_{12}}{\partial x} - \frac{1}{2} \frac{\partial M_{11}}{\partial y}\right)}{y = \bar{r}_y} \\ 
& \hspace{2.7cm} + \frac{1}{2} \bar B_2(x) \at{\frac{\partial M_{11}}{\partial x}}{y = \bar{r}_y},  \\
\gamma(p_m, p_s, x) &= -\bar B_1(x) \bar G_2(p_m, p_s ,x) + \bar B_2(x) \bar G_1(p_m, p_s, x).
\end{split}
\end{equation}
We make some relevant observations concerning the structure of the zero dynamics \eqref{eq:zero_dyn_alpha_beta_gamma}.
\begin{itemize}
\item The differential equation \eqref{eq:zero_dyn_alpha_beta_gamma} is a second order nonlinear dynamics whose stability properties are not guaranteed a priori.
In Section \ref{sec:input_en_min}, we show how it can be ensured that, by proper selection of the mass distribution parameters $p_m$, the resulting trajectories are periodic.
\item Rewrite system \eqref{eq:zero_dyn_alpha_beta_gamma} as $\ddot{x} = \psi(x, \dot{x})$.
By direct inspection of \eqref{eq:zero_dyn_alpha_beta_gamma}, we see that $\psi(x, \dot{x}) = \psi(x, -\dot{x})$.
Consider $(x, z) = (x, -\dot{x})$ and note that $dx/d(-t) = z$, $dz/d(-t) = \psi(x, z)$.
It follows that for any solution of system \eqref{eq:zero_dyn_alpha_beta_gamma} $(\bar{x}(t), \dot{\bar{x}}(t))$ such that $\dot{\bar{x}}(t_0) = 0$, for some $t_0 \in \mathbb{R}$, the forward and backward evolutions are related through $(\bar{x}(t_0 + t), \dot{\bar{x}}(t_0 + t)) = (\bar{x}(t_0 - t), -\dot{\bar{x}}(t_0 - t))$, i.e., the orbit is symmetric w.r.t. the $x$ axis.
This result implies that the target orbit associated with $(r_x(t), \dot{r}_x(t))$ must satisfy the same symmetry condition to guarantee that $r(t)$ is compatible with the zero dynamics.
\end{itemize}
\begin{assumption}\label{hyp:symmetric_reference}
For all $t_0 \in \mathbb{R}$ such that $\dot{r}_x(t_0) = 0$, it holds:
\begin{equation}
\begin{pmatrix}
r_x (t_0 + t) \\
\dot{r}_x (t_0 + t)
\end{pmatrix} =  
\begin{pmatrix}
r_x (t_0 - t) \\
- \dot{r}_x (t_0 - t)
\end{pmatrix}, \qquad \forall t \in \mathbb{R}.
\end{equation}
\end{assumption}
\begin{itemize}
\item Finally, if we exploit the explicit value of $\bar G_1$ and $\bar G_2$ in \eqref{eq:G1_G2}, the following relation between $\gamma(\cdot)$ and $S(\cdot)$ arises: 
\begin{equation}
\label{eq:gamma_S}
\gamma(p_m,p_s,x) = \zeta_S(x) S(p_s,\eta(x,\bar r_y)) + \zeta_U(p_m, x),
\end{equation}
where
\begin{equation}
\begin{split}
\zeta_S(x) &= \bar B_2 \at{\frac{\partial q_4}{\partial x}}{y = \bar{r}_y} - \bar B_1 \at{\frac{\partial q_4}{\partial y}}{y = \bar{r}_y}, \\
\zeta_U(p_m, x) &= \bar B_2 \at{\frac{\partial U_\text{g}}{\partial x}}{y = \bar{r}_y} - \bar B_1 \at{\frac{\partial U_\text{g}}{\partial y}}{y = \bar{r}_y}.
\end{split}
\end{equation}
Eq. \eqref{eq:gamma_S} indicates that system \eqref{eq:zero_dyn_alpha_beta_gamma} can be shaped via an appropriate selection of the spring $S(\cdot)$.
\end{itemize}

\section{Two-step solution}\label{sec:two-step_sol}
The approach that we propose is based on dividing the problem into two sub-problems that we solve sequentially:
\begin{enumerate}
\item Firstly, we optimize the parameter vector $p_m$ to minimize the RMS value of the linearizing torque along a given trajectory $(x(t), \dot{x}(t))$.
In particular, we constrain $(x(t), \dot{x}(t))$, solution of the zero dynamics \eqref{eq:zero_dyn_alpha_beta_gamma} with initial conditions $(r_x(0), \dot{r}_x(0))$, to be periodic and equal to $(r_x(t), \dot{r}_x(t))$.
To do so, we suppose that $\gamma$ in \eqref{eq:alpha_beta_gamma} can be shaped arbitrarily as any desired map $\hat \gamma(p_m,r_x)$, related to an ideal spring characteristic $\sigma^*(\cdot)$. 
This optimization procedure and the explicit computation of $\hat \gamma(\cdot)$ and $\sigma^*(\cdot)$ are described in detail in subsection \ref{sec:input_en_min}. 
\item Secondly, we choose a convenient class of functions $\mathcal{S}$, parametrized through $p_s$, to approximate a generic spring characteristic. 
This way, we can optimize $p_s$ so that the resulting $S(\cdot)$ approximates $\sigma^*(\cdot)$ as much as possible. 
Notably, our approach follows the intuition that a higher dimension $d$ for the parameter space leads to an improved approximation of $\sigma^*(\cdot)$. 
Subsection \ref{sec:stiff_approx} is dedicated to an in-depth presentation of this approximation strategy. 
\end{enumerate}

\subsection{Input RMS minimization}\label{sec:input_en_min}
To set up the optimization procedure, we rely on the model derived in Sections \ref{sec:min_form} and \ref{sec:feedback_lin}.
However, in place of the spring characteristic $S(p_s, \theta)$, we consider here an arbitrary function of $\theta$ (computed on the target trajectory, i.e., $\theta = \eta(r_x, \bar{r}_y)$).
Due to Assumption \ref{hyp:symmetric_reference}, for all $t \in \mathbb{R}$, we can write $\dot{r}_x^2(t) = \rho_d(r_x(t))$, $\ddot{r}_{x} = \rho_{dd}(r_x(t))$, for some functions $\rho_d(\cdot)$, $\rho_{dd}(\cdot)$.
It follows that, for any $p_m \in \mathcal{P}_m$, we can invert \eqref{eq:zero_dyn_alpha_beta_gamma} to derive an ideal term $\hat{\gamma}(p_m, r_x)$ such that $(r_x(t), \dot{r}_x(t))$ is exactly generated by the resulting zero dynamics:
\begin{equation}
\label{eq:gamma_hat}
\hat \gamma (p_m, r_x) \hspace{-1pt} = \hspace{-1pt} - \alpha(p_m, r_x) \rho_{dd}(r_x) - \beta(p_m, r_x) \rho_{d}(r_x),
\end{equation}
for all $r_x \in [r_{x_{\min}}, r_{x_{\max}}]$, where
\begin{equation}
r_{x_{\min}} = \min_{\tau \in [0, T_r)} r_x(\tau), \quad r_{x_{\max}} = \max_{\tau \in [0, T_r)} r_x(\tau). 
\end{equation}
In view of \eqref{eq:gamma_S}, we consider the following assumption.
\begin{assumption}
For all $r_x \hspace{-1pt} \in \hspace{-1pt} [r_{x_{\min}}, r_{x_{\max}}]$, it holds $\zeta_S(r_x) \neq 0$.
\end{assumption}
As a consequence, the ideal spring characteristic $\sigma(\cdot)$ required to force $x(t) = r_x(t)$ is given, for all $r_x \in [r_{x_{\min}}, r_{x_{\max}}]$, by 
\begin{equation}
\label{eq:S_hat}
\hspace{-0pt} \sigma(p_m, \eta(r_x, \bar r_y)) = \frac{\hat \gamma(p_m, r_x) - \zeta_U(p_m, r_x) }{\zeta_S(r_x)}.
\end{equation}
The input-output linearizing torque becomes:
\begin{equation}\label{eq:nu}
\begin{split}
&\nu(p_m, r_x, \dot{r}_x) \hspace{-2pt} = \hspace{-2pt} \at{\frac{\bar M_{12}(\bar C_1 \hspace{-2pt} + \hspace{-1pt} \hat{\bar G}_1) \hspace{-2pt} - \hspace{-1pt} \bar M_{11}(\bar C_2 \hspace{-2pt} + \hspace{-1pt} \hat{\bar G}_2)}{\bar M_{12}\bar B_1 - \bar M_{11}\bar B_2}}{\scriptsize\begin{matrix}x = r_x\\ \dot{x} = \dot{r}_x\end{matrix}} \\
& \hat{\bar{G}}_1(p_m,r_x) \hspace{-2pt} = \hspace{-1pt} \sigma(p_m, \eta(r_x, \bar r_y)) \hspace{-3pt} \at{\frac{\partial q_4}{\partial x}}{\scriptsize\begin{matrix}x = r_x\\ y = \bar{r}_y\end{matrix}} \hspace{-4pt} + \hspace{-3pt} \at{\frac{\partial U_\text{g}}{\partial x}}{\scriptsize\begin{matrix}x = r_x\\ y = \bar{r}_y\end{matrix}}\\
&\hat{\bar{G}}_2(p_m,r_x) \hspace{-2pt} = \hspace{-1pt} \sigma(p_m, \eta(r_x, \bar r_y)) \hspace{-3pt} \at{\frac{\partial q_4}{\partial y}}{\scriptsize\begin{matrix}x = r_x\\ y = \bar{r}_y\end{matrix}} \hspace{-4pt} + \hspace{-3pt} \at{\frac{\partial U_\text{g}}{\partial y}}{\scriptsize\begin{matrix}x = r_x\\ y = \bar{r}_y\end{matrix}} 
\end{split}
\end{equation}
where, in particular, we redefined the contributions due to the potential energy by letting $S(p_s,\eta(r_x, \bar r_y))$ be coincident with $\sigma(p_m, \eta(r_x, \bar r_y))$, for all $r_x \in [r_{x_{\min}}, r_{x_{\max}}]$.
Hence, we propose to set up an optimization problem that searches for the optimal $p_m$ (indicated with $\bar p_m$) by minimizing the RMS value of signal $\nu$. 
Besides $\bar p_m$, the optimization procedure provides a second output given by the ideal spring needed for perfect tracking $\sigma^*(\eta(r_x,\bar r_y)) = \sigma(\bar p_m, \eta(r_x,\bar r_y))$.
\par Before formalizing the optimization problem, we need to ensure that the generated trajectory is periodic and remains periodic in a neighborhood $\mathcal{N}$ of $r$.
We exploit \cite{shiriaev2006periodic}, where it is shown that if the zero dynamics \eqref{eq:zero_dyn_alpha_beta_gamma}, linearized around an equilibrium point, presents a center, then the resulting trajectories near that equilibrium point are periodic. 
We begin by imposing the following feasibility assumption. 
\begin{assumption}
\label{hyp:4}
There exists a compact set $\hat{\mathcal{P}}_m \subset \mathcal{P}_m$ such that, for all $p_m \in \hat{\mathcal{P}}_m$ there exist a point  $r_{x_0}(p_m) \in [r_{x_{\min}}, r_{x_{\max}}]$, satisfying $(r_{x_0}(p_m), \bar r_y) \in \mathcal{N}$, and such that $\hat \gamma(p_m,r_{x_0}(p_m)) = 0$.
\end{assumption} 
This way, we are able to guarantee the existence of at least an equilibrium point for the zero dynamics.
From the continuity of functions $\alpha, \beta, \hat \gamma$, the hypotheses of \cite[Theorem 3]{shiriaev2006periodic} hold, thus we ensure that the zero dynamics, linearized around $(r_{x_0}, 0)$, have a center if $\Omega(p_m) > 0$, where: 
\begin{equation}
\label{eq:Omega_cont}
\Omega(p_m) = \frac{\partial}{\partial r_x} \at{\left( \frac{\hat \gamma(p_m,r_x)}{\alpha(p_m,r_x)} \right)}{r_x = r_{x_0}(p_m)}.
\end{equation}
Therefore, to guarantee the presence of periodic orbits near $r_{x_0}$, we impose $\Omega(p_m) > 0$ as an inequality constraint in the optimization problem. 
The last requirement is the existence of at least a $p_m$ guaranteeing $\Omega(p_m) > 0$.
\begin{assumption}
There exists $\hat p_m \in \hat{\mathcal{P}}_m$ such that, for an equilibrium of the form $r_{x_0}(\hat p_m)$, it holds $\Omega(\hat p_m) > 0$.
\end{assumption}
We can thus formalize the optimization problem as follows: 
\begin{equation}\label{eq:opt_mass_cont_time}
\begin{split}
\min_{p_m \in \mathcal{P}_m} &\sqrt{\frac{1}{T_r}  \int_0^{T_r} \nu^2(p_m, r_x(t), \dot{r}_x(t)) \ dt} \\
\text{subj. to: } & \quad \eqref{eq:gamma_hat}, \quad \eqref{eq:S_hat}, \quad \eqref{eq:nu}, \quad \Omega(p_m) > 0.
\end{split}
\end{equation}
For the numerical solution of \eqref{eq:opt_mass_cont_time} we used a genetic algorithm due to the non-convex structure of the problem.
In fact, a gradient descent algorithm might converge to a local minimum potentially far away from the global one. 
Concerning the implementation of \eqref{eq:opt_mass_cont_time}, we decided to discretize the problem, i.e., we sampled the reference trajectory in $N$ points equally spaced in time (with sample time $T_{\text{s}} = T_r/N$) and we imposed the system dynamics to coincide with the reference only for the discrete trajectory samples, i.e., $(x(i T_{\text{s}}), \dot x(i T_{\text{s}}), y(i T_{\text{s}}), \dot y(i T_{\text{s}})) = (r_x(i T_{\text{s}}), \dot r_x(i T_{\text{s}}), \bar r_y, 0)$ with $i \in \{0, 1, \hdots, N-1\}$. The resulting discretized problem is the following: \\
\begin{equation}\label{eq:opt_mass_discr_time}
\begin{split}
\min_{p_m \in \mathcal{P}_m} &\sqrt{\frac{1}{N} \sum_{i=0}^{N-1} \nu^2(p_m,r_x(i T_{\text{s}}), \dot r_x(i T_{\text{s}}))} \\
\text{subj. to: } & \quad \eqref{eq:gamma_hat}, \quad \eqref{eq:S_hat}, \quad \eqref{eq:nu}, \quad \Omega_{\text{s}}(p_m) > 0.
\end{split}
\end{equation}
In this case, the constraints \eqref{eq:gamma_hat}, \eqref{eq:S_hat}, \eqref{eq:nu} are not evaluated for any $t \in [0,T_r]$ but only in correspondence of the discrete-time samples $iT_{\text{s}}$.
Moreover, for the inequality constraint, a finite difference approximation of the exact derivative defined in \eqref{eq:Omega_cont} has been introduced:
\begin{equation}
\label{eq:Omega_discr}
\Omega_{\text{s}}(p_m) = \frac{1}{r_{x2} - r_{x1}} \left( \frac{\hat \gamma(p_m,r_{x2})}{\alpha(p_m,r_{x2})} - \frac{\hat \gamma(p_m,r_{x1})}{\alpha(p_m,r_{x1})}\right),
\end{equation}
where $r_{x1} = r_x(\bar \imath T_{\text{s}})$ and $r_{x2} = r_x((\bar \imath + 1) T_{\text{s}})$, for some $\bar \imath \in \{0, \hdots, N - 1\}$ such that:
\begin{equation}
\frac{\hat \gamma(p_m,r_{x1})}{\alpha(p_m,r_{x1})} \cdot \frac{\hat \gamma(p_m,r_{x2})}{\alpha(p_m,r_{x2})} < 0.
\end{equation}

\subsection{Nonlinear stiffness approximation}\label{sec:stiff_approx}
Consider the ideal spring characteristic, $\sigma^*(\cdot)$, obtained in the previous subsection. 
If we were able to realize exactly such characteristic, then the term $\gamma$ of the zero dynamics \eqref{eq:zero_dyn_alpha_beta_gamma} would match the desired profile, so that \eqref{eq:gamma_hat} would hold true (at the discrete samples of the reference trajectory), forcing $(x(t), \dot{x}(t))$ to be (almost) equal to $(r_x(t), \dot{r}_x(t))$. 
The purpose of this subsection is twofold: firstly, we present a strategy to parameterize through a vector $p_s$ the characteristic of a nonlinear spring; then, we set up an optimization problem that searches the best parameters, $\bar{p}_s$, that approximate $\sigma^*(\theta)$, for all $\theta \in [\theta_{\min}, \theta_{\max}]$, where
\begin{equation}\label{eq:theta_min_theta_max}
\theta_{\min} = \min_{\tau \in [0, T_r)} \eta(r_x(\tau), \bar{r}_y), \quad \theta_{\max} = \max_{\tau \in [0, T_r)} \eta(r_x(\tau), \bar{r}_y). 
\end{equation}
\par There are several works in the literature related to the design of nonlinear springs (see \cite{friswell2012vibration} and references therein).
In this context, particularly interesting are the approaches found in \cite{radaelli2017static} and \cite{yao2018novel}, where a piecewise linear stiffness is used to fit the desired nonlinear characteristic.
As a consequence, it is possible to physically implement the computed stiffness through a combination of multiple linear springs.
This procedure has been employed, e.g., in \cite{radaelli2017static} to compensate the weight of an inverted pendulum in a wide interval of angular configurations.
\par Taking advantage of these results, we propose to approximate $\sigma^*(\cdot)$ with a piecewise linear stiffness.
To parameterize such a profile, we suppose that the overall characteristic is given by the sum of different elementary components.
In particular, we start the approximation with a linear spring with stiffness $k_0$ and rest angle $\theta_0$.
Then, we increase the complexity of the characteristic by introducing pairs of positive-negative sub-springs.
By positive sub-spring, we intend a spring that gives no contribution until $\theta$ reaches a specific threshold $\theta^{+}$.
For $\theta \geq \theta^{+}$, the positive sub-spring has a linear characteristic of slope $k^{+}$.
Conversely, a negative sub-spring is a linear spring (with stiffness $k^{-}$) for $\theta \leq \theta^{-}$, for a given threshold $\theta^{-}$.
In view of the possibility of realizing springs with negative stiffness \cite{lee2007design}, we allow $k^{+}$ and $k^{-}$ to be either positive or negative.
On the other hand, we impose $k_0 \geq 0$ for the linear spring.
\par The overall characteristic is thus obtained through the combination of one linear spring and $n$ pairs of positive-negative sub-springs.
Notice that increasing $n$ improves the characteristic approximation, even though it increases the complexity in the actual realization of the spring.
Based on the described modeling strategy, the ensuing analytical representation of the spring characteristic is: 
\begin{equation}
S(p_s, \theta) = S_0(k_0, \theta_0, \theta) + \sum_{j=1}^{n} S^{+}_j (k^{+}_j \hspace{-2pt}, \theta^{+}_j \hspace{-2pt}, \theta) + S^{-}_j(k^{-}_j \hspace{-2pt}, \theta^{-}_j \hspace{-2pt}, \theta) 
\end{equation}
where
\begin{equation}
\begin{split}
S_0(k_0, \theta_0, \theta) &= k_0(\theta - \theta_0); \\
S^{+}_j(k^{+}_j \hspace{-2pt}, \theta^{+}_j \hspace{-2pt}, \theta) &= \begin{cases}
0, \quad \hspace{-3pt}  &\text{if} \ \theta < \theta^{+}_j \\
k^{+}_j(\theta - \theta^{+}_j), \; \hspace{-3pt} &\text{if} \ \theta \geq \theta^{+}_j
\end{cases} \quad \hspace{-3pt} j \in\{1, \hdots, n\},\\
S^{-}_j(k^{-}_j \hspace{-2pt}, \theta^{-}_j \hspace{-2pt}, \theta) &= \begin{cases}
k^{-}_j(\theta - \theta^{-}_j), \; \hspace{-3pt} &\text{if} \ \theta \leq \theta^{-}_j \\
0, \quad \hspace{-3pt}  &\text{if} \ \theta > \theta^{-}_j 
\end{cases} \quad \hspace{-3pt} j \in \{1, \hdots, n\}.
\end{split}
\end{equation}
We can thus introduce an optimization algorithm aiming to approximate $S^*(\theta)$. We collect all the parameters determining the spring characteristic in vector $p_s$: 
\begin{equation}
p_s = (k_0 \ \theta_0 \ k^{+}_1 \ \theta^{+}_1 \ k^{-}_1 \ \theta^{-}_1 \ \hdots \ k^{+}_n \ \theta^{+}_n \ k^{-}_n \ \theta^{-}_n)^\top \in \mathbb{R}^d
\end{equation}
where $d = 2 + 4n$, since every pair of positive-negative sub-springs introduces four parameters ($k^{+}, \theta^+, k^{-}, \theta^-$), while the linear spring is characterized by two parameters ($k_0, \theta_0$).
The domain of $p_s$ is $\mathcal{P}_s = [0, \bar k] \times [\theta_{\min}, \theta_{\max}] \times \{[-\bar k, \bar k] \times [\theta_{\min}, \theta_{\max}]\}^{2n}$, where $\bar k$ is the maximum slope (in absolute value) admissible for the linear characteristic of any spring or sub-spring, while $\theta_{\min}$ and $\theta_{\max}$ are defined in \eqref{eq:theta_min_theta_max}.
Therefore, we look for an optimal parameter configuration, $\bar p_s$, minimizing the mismatch between the actual spring characteristic and the ideal one, both evaluated in $N$ samples of angle $\theta = \eta(r_x(i T_{\text{s}}), \bar r_y)$ corresponding to the $N$ samples of $r_x$ imposed in the mass distribution optimization problem: 
\begin{equation}\label{eq:opt_spring}
\min_{p_s \in \mathcal{P}_s} \frac{1}{N} \sum_{i=0}^{N-1} [S(p_s, \eta(r_x(i T_{\text{s}}), \bar r_y)) - \sigma^*(\eta(r_x(i T_{\text{s}}), \bar r_y))]^2.
\end{equation}
As for the mass distribution optimization, to account for the non-convex cost function, we employed the genetic algorithm to solve the problem. \\
\begin{remark}
In this second optimization problem, we have not imposed any condition regarding the properties of the equilibrium point of the resulting zero dynamics.
In fact, we assumed implicitly that the obtained $\gamma(\bar p_m, \bar p_s, r_x)$ is sufficiently close to $\hat \gamma(\bar p_m, r_x)$.
Hence, both the equilibrium of the zero dynamics and the trajectories near such equilibrium are expected to have similar properties w.r.t. the ideal ones.
If this were not the case, we could always increase the accuracy in the spring characteristic approximation by increasing the number of sub-springs, so that a better approximation of $\hat \gamma(\bar p_m, r_x)$ is obtained. 
\end{remark}

\subsection{Overall problem solution}
We can sum up our solution to Problem \ref{prob1}.
Firstly, from the mass distribution optimization we obtain the optimal $\bar p_m$ and the ideal spring characteristic $\sigma^*(\cdot)$.
Secondly, choosing the number $d = 2 + 4n$ of parameters used to model the spring profile, we run the spring parameter optimization determining $\bar{p}_s$.
This second step can be repeated more than once, with different values of $n$, i.e., with a different degree of accuracy in the approximation of $\sigma^*(\cdot)$.
Finally, the input torque that completes the solution of the problem is the following: 
\begin{equation}
\bar u(\bar p_m, \bar p_s, q, \dot q) = - \frac{f_y(\bar p_m, \bar p_s, x(q), \dot{x}(q, \dot{q}), y(q), \dot{y}(q, \dot{q}))}{g_y(\bar p_m, x(q), y(q))}
\end{equation}
with $f_y$ and $g_y$ defined in \eqref{eq_f_g_explicit}, $(x\; y)^\top = h(q)$, $(\dot{x}\; \dot{y})^\top = \frac{\partial h}{\partial q}(q) \dot{q}$ and $h$ as in \eqref{eq:direct_kin}.
Recalling Remark \ref{rem1}, we highlight that this input action does not stabilize the system on the reference trajectory.
Instead, $\bar u$ is the feedback-linearizing input torque that keeps the system on the reference $r$ in perfect tracking conditions.

\section{Numerical results}\label{sec:num_results}

\begin{table}
\begin{center}
\caption{Parameters of the Mechanism Links}\label{tab:link_parameters}
\addtolength{\tabcolsep}{-4pt}  
\begin{tabular}{lr | lr | lr}\hline
{\scriptsize Length} & {\scriptsize [$\text{m}$]} & {\scriptsize Total mass} & {\scriptsize [$\text{kg}$]} & {\scriptsize Moment of inertia} & {\scriptsize[$\text{kg m}^2$]} \\ 
\hline
{\scriptsize $l_1$}  &{\scriptsize $0.080$} & {\scriptsize $m_1$} & {\scriptsize $0.071$} & {\scriptsize $J_1$} & {\scriptsize $0.747\times10^{-4}$}\\ 
{\scriptsize $l_2$}  &{\scriptsize $0.235$} & {\scriptsize $m_2$} & {\scriptsize $0.195$} & {\scriptsize $J_2$} & {\scriptsize $10.413\times10^{-4}$}\\
{\scriptsize $l_3$}  &{\scriptsize $0.052$} & {\scriptsize $m_3$} & {\scriptsize $0.049$} & {\scriptsize $J_3$} & {\scriptsize $0.345\times10^{-4}$}\\
{\scriptsize $l_4$}  &{\scriptsize $0.135$} & {\scriptsize $m_4$} & {\scriptsize $0.115$} & {\scriptsize $J_4$} & {\scriptsize $2.430\times10^{-4}$}\\ 
\hline
\end{tabular}
\addtolength{\tabcolsep}{4pt}
		
\end{center}
	
\end{table}

\begin{figure}[t]
	
\psfragscanon
	
\psfrag{x} [B][B][0.7][0]{(a) time [s]}
\psfrag{y} [B][B][0.7][0]{$r_x, \bar x$ [m]}
\includegraphics[clip = true, width = 0.23\textwidth]{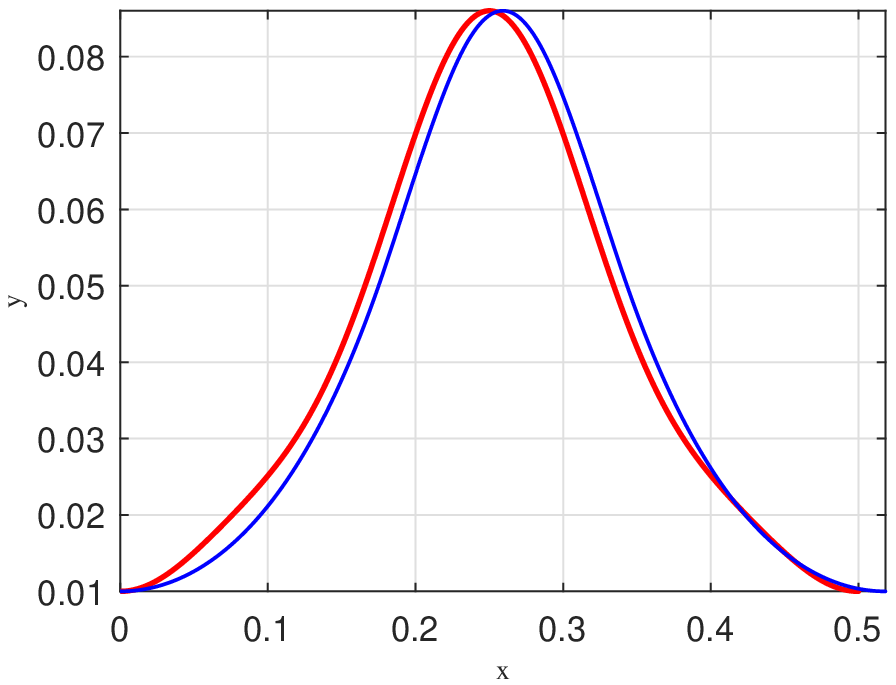}
\hspace{0.001\textwidth}
\psfrag{x} [B][B][0.7][0]{(d) time [s]}
\psfrag{y} [B][B][0.7][0]{$r_x, \bar x$ [m]}
\includegraphics[clip = true, width = 0.23\textwidth]{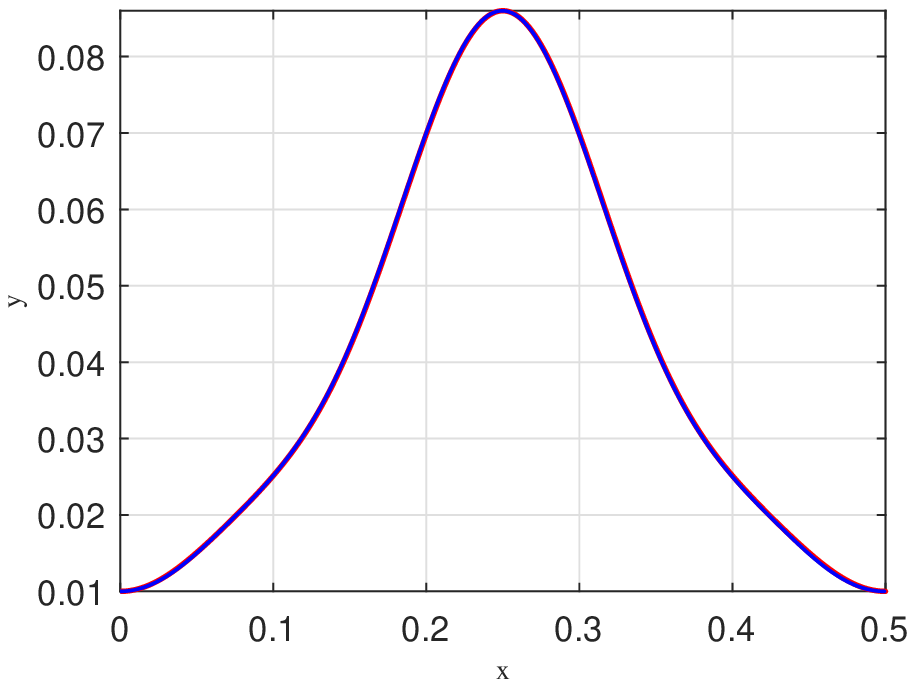}
	
\vspace{2pt}
	
\psfrag{x} [B][B][0.7][0]{(b) time [s]}
\psfrag{y} [B][B][0.7][0]{$\dot r_x, \dot{\bar x}$ [m/s]}
\includegraphics[clip = true, width = 0.23\textwidth]{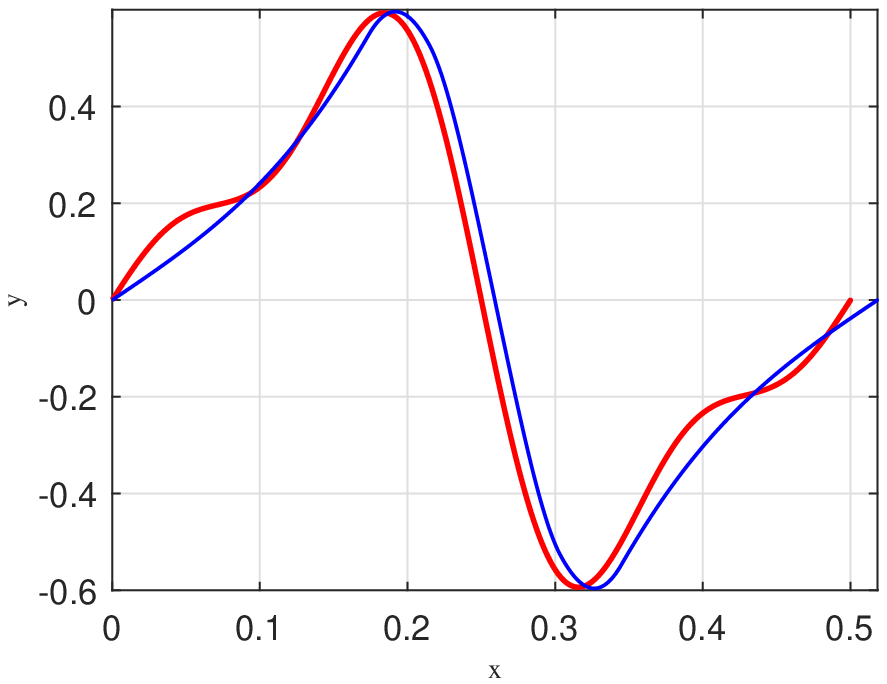}
\hspace{0.001\textwidth}
\psfrag{x} [B][B][0.7][0]{(e) $\theta$ [rad]}
\psfrag{y} [B][B][0.7][0]{$\dot r_x, \dot{\bar x}$ [m/s]}
\includegraphics[clip = true, width = 0.23\textwidth]{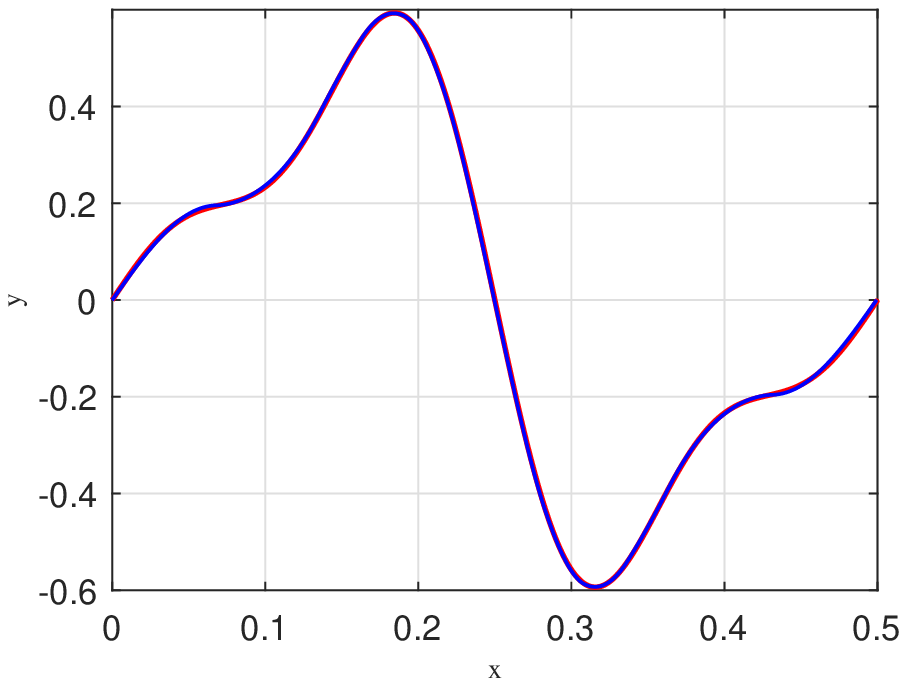}
	
\vspace{2pt}
	
\psfrag{x} [B][B][0.7][0]{(c) $r_x, \bar x$ [m]}
\psfrag{y} [B][B][0.7][0]{$\dot r_x, \dot{\bar x}$ [m/s]}
\includegraphics[clip = true, width = 0.23\textwidth]{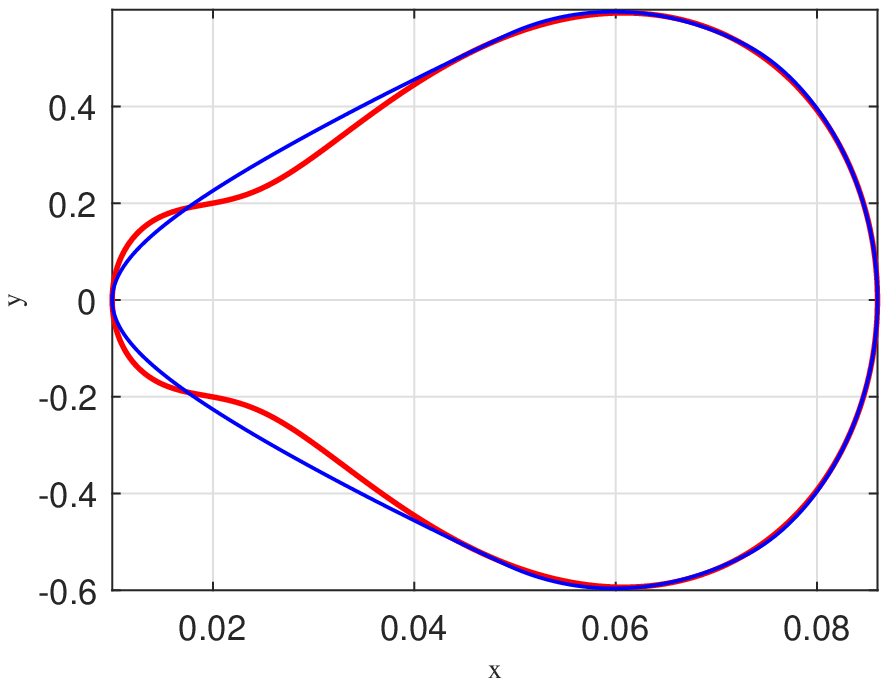}
\hspace{0.001\textwidth}
\psfrag{x} [B][B][0.7][0]{(f) $r_x, \bar x$ [m]}
\psfrag{y} [B][B][0.7][0]{$\dot r_x, \dot{\bar x}$ [m/s]}
\includegraphics[clip = true, width = 0.23\textwidth]{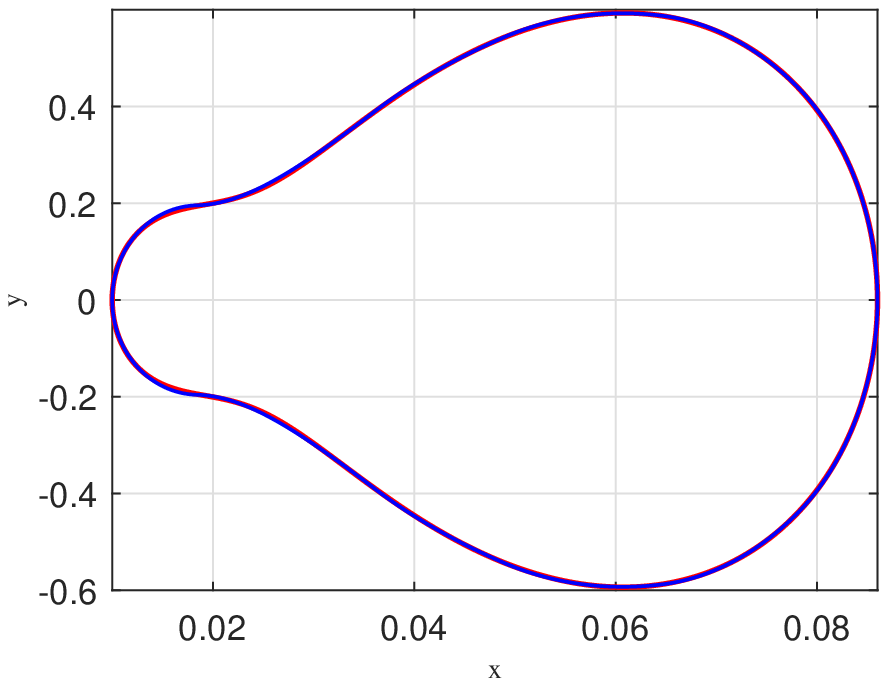}

\caption{Periodic motion of the zero dynamics for the first reference trajectory. (a): reference $r_x$ (red) and zero dynamics evolution $\bar x$ (blue), $n=1$. (b): reference $\dot r_x$ (red) and zero dynamics evolution $\dot{\bar x}$ (blue), $n=1$. (c) target orbit (red) and the one generated by the zero dynamics (blue), $n=1$. (d): reference $r_x$ (red) and zero dynamics evolution $\bar x$ (blue), $n=3$. (e): reference $\dot r_x$ (red) and zero dynamics evolution $\dot{\bar x}$ (blue), $n=3$. (f) target orbit (red) and the one generated by the zero dynamics (blue), $n=3$.}
	
\label{fig:sim_ref1}
\end{figure}

\begin{figure}
	
\psfragscanon
	
\psfrag{x} [B][B][0.7][0]{(a) $\theta$ [rad]}
\psfrag{y} [B][B][0.7][0]{$\sigma^*(\theta), S(\bar p_s, \theta)$ [Nm]}
\includegraphics[clip = true, width = 0.23\textwidth]{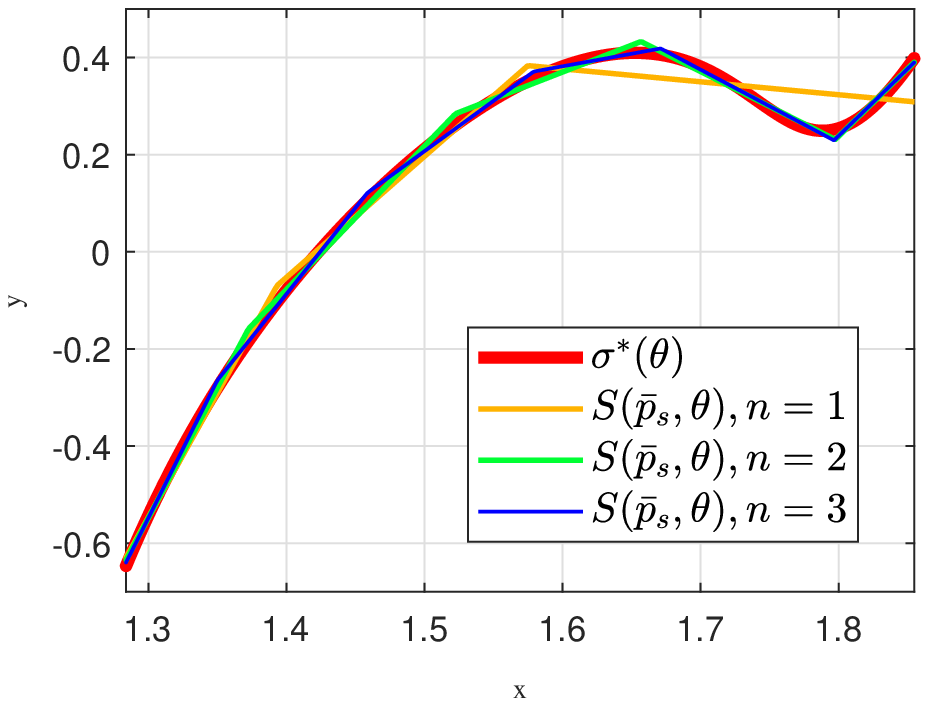}
\hspace{0.001\textwidth}
\psfrag{x} [B][B][0.7][0]{(b) $\theta$ [rad]}
\psfrag{y} [B][B][0.7][0]{$\sigma^*(\theta) - S(\bar p_s, \theta)$ [Nm]}
\includegraphics[clip = true, width = 0.23\textwidth]{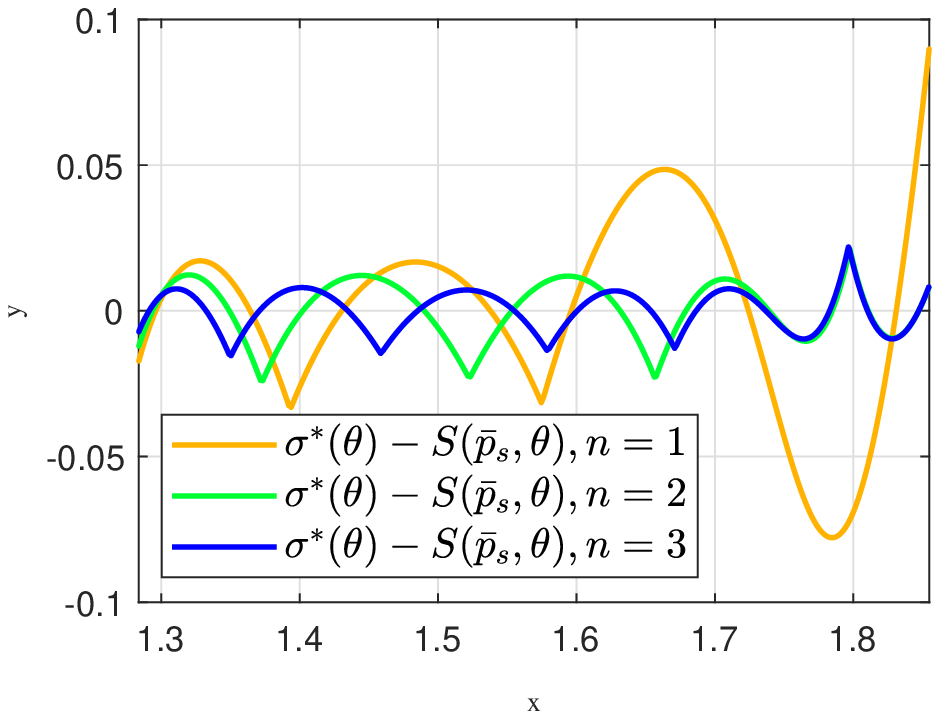}
	
\caption{Spring optimization results for the first reference trajectory. (a): $\sigma^*(\theta)$ (red) and $S(\bar p_s,\theta)$ comparison with $n \in \{1,2,3\}$. (b):  $\sigma^*(\theta) - S(\bar p_s,\theta)$ with $n \in \{1, 2, 3\}$.}
	
\label{fig:spring_approx_ref1}
\end{figure}

In this section, we present the results of our optimization strategy applied to the mechanism of Fig. \ref{fig:mechanism}.
The parameters of the links are summarized in Table \ref{tab:link_parameters}.
In particular, the values of the masses and the moments of inertia (computed w.r.t. the center of mass of each link) refer to the links without the additional masses.
Such additional masses were allowed to be at most equal to $0.1$kg.
The domain considered in the mass distribution optimization problem is given by the following box: $\mathcal{P}_m = [0,0.1]^2 \times [-0.05,0.05]^2$.
For the domain $\mathcal{P}_s$, described in Section \ref{sec:stiff_approx}, we chose $\bar k = 10$Nm/rad, while $\theta_{\min}$ and $\theta_{\max}$ have been computed as in \eqref{eq:theta_min_theta_max} according to the assigned reference trajectory.
\par Concerning the solution of problems \eqref{eq:opt_mass_discr_time} and \eqref{eq:opt_spring}, we ran the optimization on an AMD Ryzen Threadripper 2950X 16-core processor with $3.5$GHz base clock frequency and $4.4$GHz maximum boost frequency.
In both optimization steps, we fixed $N = 1000$ and we employed the genetic algorithm available in Matlab, \texttt{ga}.
\par Two different horizontal reference trajectories $r_x(t)$ have been considered for testing our two-step solution.
Both these trajectories satisfy Assumption \ref{hyp:symmetric_reference} and have a period $T_r$ equal to $0.5$s.
In both cases, the end-effector height $\bar r_y$ is set to $0.15$m, while the position of joint $\mathcal{J}_1$ is fixed to $(-0.19, 0.15)$m in the $(e_x,e_y)$ frame.
\par The first horizontal trajectory taken into account has the profiles of $r_x(t)$ and $\dot r_x(t)$ shown in red respectively in Figs. \ref{fig:sim_ref1}-(a) and \ref{fig:sim_ref1}-(b), while the red plot of Fig. \ref{fig:sim_ref1}-(c) represents the target orbit associated with $(r_x(t),\dot r_x(t))$.
The forward motion of this trajectory (and conversely, the backward phase) consists of an initial low-speed phase, followed by a second high-speed phase.
The optimized mass distribution parameters are found in $\bar p_m = (0.100\text{kg}, 0.062\text{kg}, -0.050\text{m}, 0.022\text{m})^\top$, providing an ideal input with RMS value equal to $0.143$Nm. Notice that, without any additional mass, the RMS value of the input would be equal to $0.193$Nm.
Therefore, a reduction of $25.8 \%$ in the RMS value of the input is obtained due to the mass distribution optimization. 
The ideal spring characteristic $\sigma^*(\theta)$, resulting from the first optimization procedure, is approximated with $1, 2$, and $3$ couples of sub-springs, i.e., $n \in \{1,2,3\}$. 
Fig. \ref{fig:spring_approx_ref1}-(a) shows the comparison between $\sigma^*(\theta)$, in red, and $S(\bar p_s, \theta)$ obtained with the three different choices of $n$ (color associations: orange corresponds to $n=1$, green corresponds to $n=2$, blue corresponds to $n=3$).
As expected, Fig. \ref{fig:spring_approx_ref1}-(b) confirms that a higher number of sub-springs improves the approximation, i.e., the the mismatch between $\sigma^*(\theta)$ and $S(\bar p_s, \theta)$ is reduced. 
Figs. \ref{fig:sim_ref1}-(a), \ref{fig:sim_ref1}-(b), \ref{fig:sim_ref1}-(c), and Figs. \ref{fig:sim_ref1}-(d), \ref{fig:sim_ref1}-(e), \ref{fig:sim_ref1}-(f) illustrate the zero dynamics behavior (in blue), compared with the reference (in red), obtained respectively in the case of $n=1$ and $n=3$.
Notably, it is possible to appreciate a significantly improved approximation of the target orbit as a consequence of the more accurate approximation of $\sigma^*(\theta)$.
\par
\begin{figure}
	
\psfragscanon
	
\psfrag{x} [B][B][0.7][0]{(a) $\theta$ [rad]}
\psfrag{y} [B][B][0.7][0]{$\sigma^*(\theta), S(\bar p_s, \theta)$ [Nm]}
\includegraphics[clip = true, width = 0.23\textwidth]{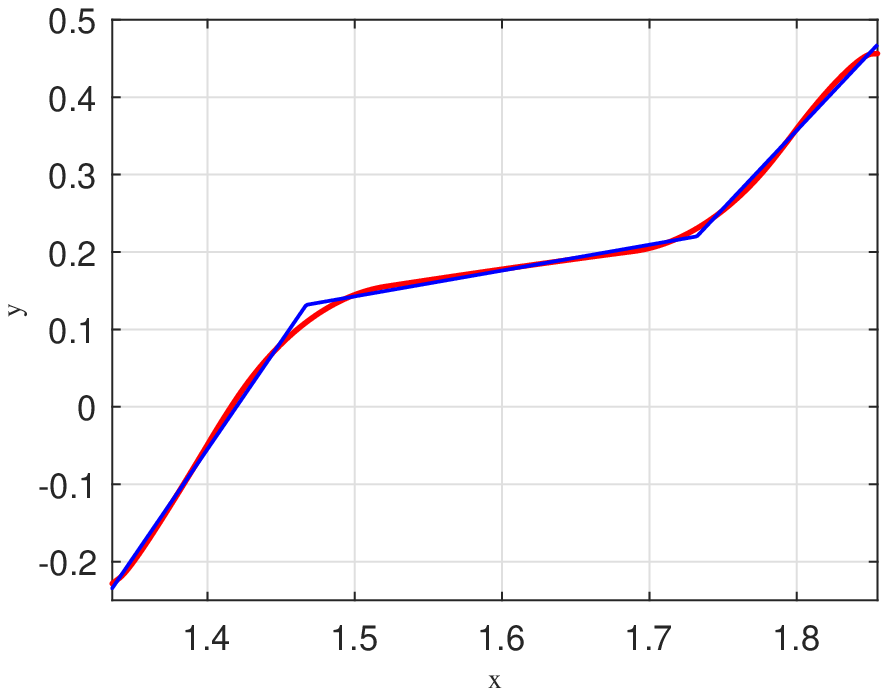}
\hspace{0.001\textwidth}
\psfrag{x} [B][B][0.7][0]{(b) time [s]}
\psfrag{y} [B][B][0.7][0]{$r_x, \bar x$ [m]}
\includegraphics[clip = true, width = 0.23\textwidth]{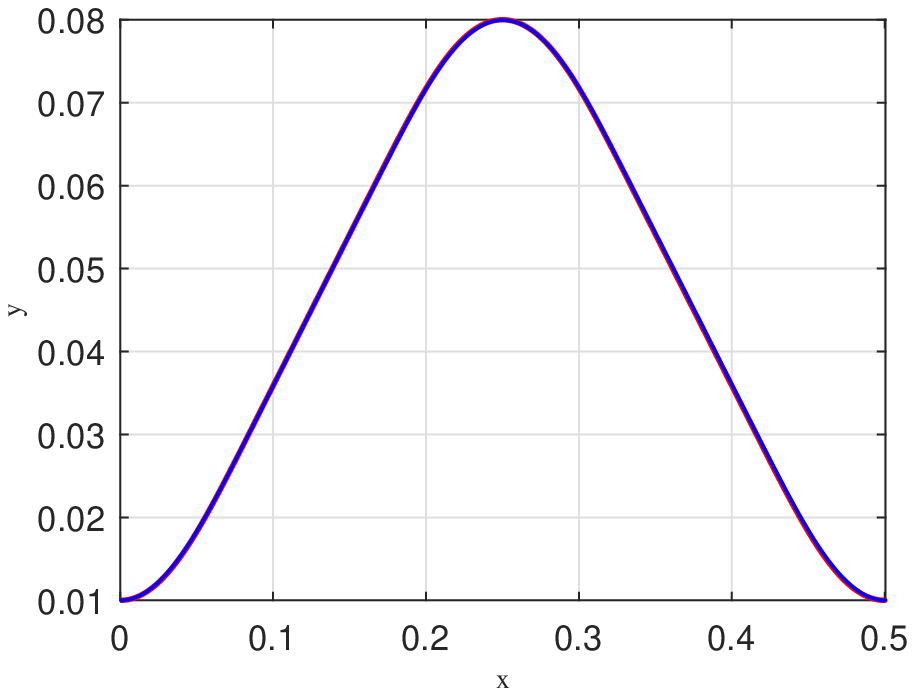}
	
\vspace{2pt}
	
\psfrag{x} [B][B][0.7][0]{(c) time [s]}
\psfrag{y} [B][B][0.7][0]{$\dot r_x, \dot{\bar x}$ [m/s]}
\includegraphics[clip = true, width = 0.23\textwidth]{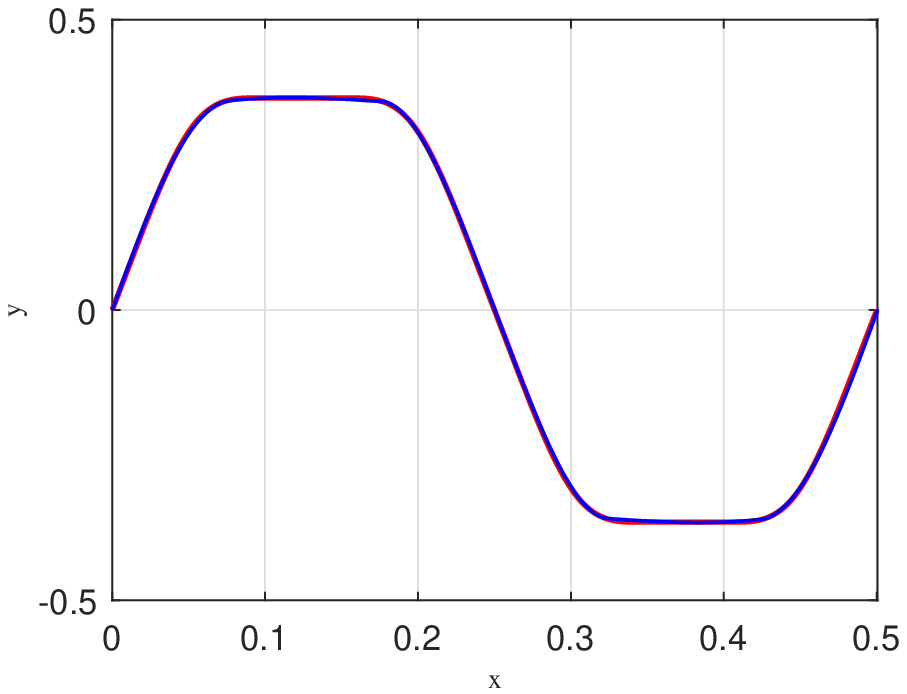}
\hspace{0.001\textwidth}
\psfrag{x} [B][B][0.7][0]{(d) $r_x, \bar x$ [m]}
\psfrag{y} [B][B][0.7][0]{$\dot r_x, \dot{\bar x}$ [m/s]}
\includegraphics[clip = true, width = 0.23\textwidth]{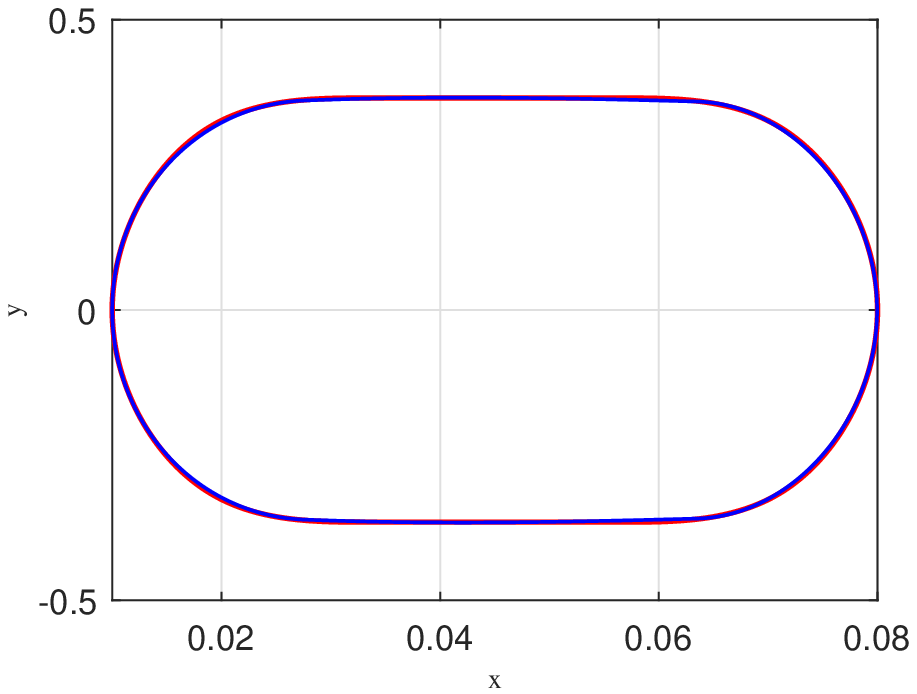}
	
\caption{ Second reference trajectory: spring characteristic approximation and periodic motion of the zero dynamics. (a): $\sigma^*(\theta)$ (red), $S(\bar p_s,\theta)$ (blue), $n=1$. (b): reference $r_x$ (red) and zero dynamics evolution $\bar x$ (blue), $n=1$. (b): reference $\dot r_x$ (red) and zero dynamics evolution $\dot{\bar x}$ (blue), $n=1$. (c) target orbit (red) and the one generated by the zero dynamics (blue), $n=1$.}
	 
\label{fig:spring_approx_sim_ref2}
\end{figure}

 The second horizontal reference trajectory that we considered has a trapezoidal velocity profile (with appropriate smoothing to satisfy the requirement $r_x \in \mathcal{C}^2$).
The optimized mass distribution parameters are found in $\bar p_m = (0.100\text{kg}, 0.044\text{kg}, -0.050\text{m}, 0.017\text{m})^\top$ and provide a reduction of $25.9 \%$ w.r.t. the RMS value of the input obtained in case of $m_{A_3}=m_{A_4}=0$kg (i.e., $0.138$Nm instead of $0.187$Nm).
In this case, we approximated $\sigma^*(\theta)$ with only one pair of sub-springs, obtaining a very satisfactory zero dynamics matching. 
Fig. \ref{fig:spring_approx_sim_ref2} reports the profiles of $S(\bar p_s,\theta)$ and $(\bar x(t), \dot{\bar x}(t))$, compared with $\sigma^*(\theta)$ and $(r_x(t), \dot r_x(t))$.

\section{Conclusions}\label{sec:conclusions}
We presented a control-oriented structural design strategy that generates an optimal periodic motion for a 2-DOF underactuated mechanism.
The proposed solution is based on the co-design of an input-output linearizing torque and some system's structural parameters.
In particular, we formalized a two-step optimization strategy.
Firstly, we minimized the RMS value of the input by selecting the mass distribution and obtaining a nonlinear stiffness that imposes the desired periodic behavior for the system's zero dynamics.
Afterward, we approximated the nonlinear spring through a piecewise linear characteristic for a simpler physical implementation of the system.
Future works will be dedicated to integrating the two optimization steps and extending the approach to more general underactuated mechanisms.

\bibliographystyle{ieeetr}
\bibliography{AIM_2021_bibliography}

\end{document}